\documentclass[12pt]{iopart}
\usepackage{iopams}  
\usepackage{cite}
\expandafter\let\csname equation*\endcsname\relax

\expandafter\let\csname endequation*\endcsname\relax
\usepackage{amsmath}
\usepackage[]{hyperref} 
 
 \newcommand{\be}{\begin{equation}}
\newcommand{\ee}{\end{equation}}

\makeatletter

\newcommand*\xbar[1]{%
  \hbox{%
    \vbox{%
      \hrule height 0.5pt 
      \kern0.3ex
      \hbox{%
        \kern-0.0em
        \ensuremath{#1}%
        \kern-0.0em
      }%
    }%
  }%
} 

\makeatother

\makeatletter
\renewcommand\theequation%
{\thesection.\arabic{equation}}
\@addtoreset{equation}{section}
\makeatother

\usepackage{xcolor}

\begin{document}

\title{The BMS group in $D= 6$ spacetime dimensions}

\author{Oscar Fuentealba$^1$ \& Marc Henneaux$^{1,2}$}

\address{$^1$ Universit\'e Libre de Bruxelles and International Solvay Institutes, ULB-Campus Plaine CP231, B-1050 Brussels, Belgium\\
$^2$ Coll\`ege de France,  Universit\'e PSL, 11 place Marcelin Berthelot, 75005 Paris, France}
\ead{oscar.fuentealba@ulb.be, marc.henneaux@ulb.be}
\vspace{10pt}
\begin{indented}
\item[] December 2023
\end{indented}

\begin{abstract}
The asymptotic structure of gravity in $D=6$ spacetime dimensions is described at spatial infinity in the asymptotically flat context through Hamiltonian (ADM) methods. Special focus is given on the  BMS supertranslation subgroup. It is known from previous studies that the BMS group contains more supertranslations as one goes from $D=4$ to $D=5$.  Indeed, while the supertranslations are described by one single function of the angles in $D=4$, four such functions are neeeded in $D=5$. We  consider the case $D=6$ with the aim of determining whether the number of supertranslations keeps increasing with the dimension or remains equal to the number found in $D=5$.   We show that even though there is apparent room for more supertranslations, their number remains equal to the $D=5$ value (four): the potentially new supertranslations turn out to define proper gauge transformations corresponding to a redundancy in the description of the system.  Critical in the analysis are the boundary conditions chosen to yield a well-defined canonical formalism. Given the computational (but not conceptual) complexity as one increases the dimension, we explicitly discuss the linearized theory and argue that asymptotically, this analysis provides the correct picture.   We conclude by considering higher spacetime dimensions where we indicate that the number of physically relevant supertranslations remains equal to four independently of the dimension $\geq 5$.

\vspace{2mm}
{\it Contribution to Stanley Deser memorial volume ``Gravity, Strings and Beyond''}
\end{abstract}

%
%
%
%
%




\section{Introduction}
Stanley Deser made remarkable contributions to the dynamical understanding of Einstein theory of gravity \cite{Arnowitt:1962hi} (see also \cite{Dirac:1958sc,Dirac:1958jc}).  He  also provided major insight on the role of surface integrals at infinity (energy, angular momentum) in the Hamiltonian formulation \cite{Arnowitt:1962hi}.  The current paper is an extension of the ideas developed in \cite{Arnowitt:1962hi} to higher dimensions - another subject dear to Stanley's heart - and directly relies on his pioneering advances.

The asymptotic symmetry group of Einstein gravity in four spacetime dimensions is the Bondi-Metzner-Sachs (BMS) group \cite{Bondi:1962px,Sachs:1962wk,Sachs:1962zza}.  This was established first at null infinity in the late 1950's - early 1960's.  It is only recently that the very same symmetry was shown to be present at spatial infinity.  This was achieved by insisting that a symmetry of a theory should be in particular a symmetry of the action (exactly and not just up to boundary terms at spatial infinity), and by providing boundary conditions that yielded a consistent variational formulation (finite action, invariance under at least the Poincar\'e group) without imposing gauge conditions that would freeze ``improper'' gauge symmetries \cite{Henneaux:2018hdj,Henneaux:2019yax} (see \cite{Benguria:1976in} for the crucial concepts of ``proper'' versus ``improper'' gauge symmetries and \cite{Henneaux:2018cst} for a BMS invariant Hamiltonian formulation with different asymptotic conditions).  We compare in \ref{App:Comparison} our approach to asymptotic symmetries at spatial infinity with other approaches yieding symmetry groups different from the BMS group.

Not only can one formulate boundary conditions at spatial infinity leading to the BMS algebra as asymptotic symmetry algebra, but one can also make the explicit asymptotic integration of the equations along the lines pursued in \cite{Fried1,Friedrich:2017cjg,Herberthson:1992gcz,Troessaert:2017jcm} to make the connection with null infinity at the ``critical surfaces''.  By so doing, one can prove that the boundary conditions of \cite{Henneaux:2018hdj}, which implements parity-restrictions in a supertranslation-invariant manner on the initial data,  eliminate the potential leading logarithmic singularity that might otherwise appear (subleading logarithmic terms will of course generically be unavoidable)\footnote{The existence of null infinity with given smoothness properties should be viewed as a dynamical question in Einstein's theory of gravity.  With generic Cauchy data, logarithmic terms will generally develop at leading orders at null infinity \cite{Christodoulou:1993uv,Christodoulou:2000}.}. In that analysis, the matching conditions of \cite{Strominger:2013jfa,Strominger:2017zoo} are consequences of the parity conditions (up to a diffeomorphism) imposed on the initial data \cite{Henneaux:2018hdj}\footnote{This result derived in 2018 in  \cite{Henneaux:2018hdj} by Hamiltonian methods has been confirmed by different methods, see e.g. \cite{Mohamed:2023jwv} for a recent paper on the subject arriving at similar conclusions.}.

The connection with null infinity has been used recently \cite{Fuentealba:2023syb} to show the equivalence of the supertranslation-invariant redefinitions of the angular momentum proposed at null infinity \cite{Mirbabayi:2016axw,Bousso:2017dny,Javadinezhad:2018urv,Javadinezhad:2022hhl,Javadinezhad:2023mtp,Compere:2019gft,Compere:2021inq,Compere:2023qoa,Chen:2021szm,Chen:2021kug,Chen:2021zmu}
and spatial infinity \cite{Fuentealba:2022xsz}.  In particular,  which coefficients of the metric encode the information about the memory effect in the asymptotic $1/r$-expansion as one goes to infinity along spacelike directions (and why this is physically meaningful) was also clarified in \cite{Fuentealba:2023syb}.

The asymptotic structure of Einstein gravity in five spacetime dimensions was analysed at spatial infinity in \cite{Fuentealba:2021yvo,Fuentealba:2022yqt}, also along Hamiltonian lines.  The main new feature was the enlargement of the supertranslation subgroup: while it depends on one function of the angles in $4$ dimensions, it depends on four such functions in $5$ dimensions.  This is because parity conditions do not need to be imposed to get a consistent Hamiltonian formulation and furthermore subleading supertranslations become ``improper'' and hence physically relevant.  The generators of the subleading 
supertranslations are canonically conjugate to the generators of ordinary supertranslations, permitting a supertranslation-invariant redefinition of the Lorentz generators \cite{Fuentealba:2023hzq}.  An interesting feature is that the Hamiltonian analysis in five spacetime dimensions at spatial infinity is a direct conceptual extension of the four-dimensional one, while null infinity seems to raise new subtleties.

The question then arises as to how this extends to higher dimensions, in particular, what is the description of the relevant BMS group in $D=6$ spacetime dimensions?  Does it keep increasing with the dimensions? Or do the supertranslations remain parametrized by four functions of the angles as in $5$ spacetime dimensions? This is the question explored in this paper.  We show that while more functions of the angles parametrize the asymptotic behaviour of the fields at spatial infinity, only four are physically relevant, i.e., define improper gauge symmetries with non-vanishing charges.  The other functions of the angles appearing in the asymptotic expansion correspond to proper gauge symmetries.  So, once the proper gauge symmetries are factored out, one gets physically relevant BMS supertranslations parametrized by four functions of the angles, as in $5$ spacetime dimensions. 

Given the technical intricacies of the theory, which grow with the number of spacetime dimensions, we consider the linearized (Pauli-Fierz) theory - which has been the starting point of many decisive works of Stanley. As shown in \cite{Fuentealba:2020ghw} for $D=4$, this theory keeps all the relevant asymptotic symmetry features of the full theory.  One can easily verify that this is also the case in $D=5$ and there is no reason to expect the pattern to be altered for higher $D$'s.   The analysis of the linearized theory sheds furthermore new light on the origin of the supertranslations:  the pure supertranslations originate from the improper gauge symmetries of the Pauli-Fierz theory, while the standard translations originate from global symmetries of the Pauli-Fierz theory, which become improper gauge symmetries upon ``gauging''.

The linearized approach was also followed by the authors of \cite{Kapec:2015vwa} in the investigation of the asymptotic properties of gravity in six and higher even spacetime dimensions at null infinity.  It was also argued there that linearization preserved the relevant asymptotic features.

Our paper is organized as follows. In Section {\bf \ref{Sec:Presentation}}, we briefly present the theory: its action, the gauge transformations and a preliminary form of the asymptotic conditions on the canonical variables.  We then show in Section {\bf \ref{FinitenessSymplectic}}  that finiteness of the symplectic structure imposes constraints on the fields appearing in the asymptotic expansion.  The same constraints are shown in Section {\bf \ref{Sec:FinitessGen}} to arise also from the requirement of finiteness of the generators of gauge transformations.  The final form of the asymptotic conditions that takes into account all the restrictions is given in Section {\bf \ref{Asymptotia1}}.  The improper gauge transformations are discussed next in Section {\bf \ref{Sec:GTFormAnd Algebra}} and, taking into account the results of Section {\bf \ref{Sec:FinitessGen}}, shown to depend on four functions of the angles (minus the $l=0$ spherical harmonics for two of them and the $l=1$ harmonics for the other two).  The Poisson bracket algebra of the canonical generators is also displayed and importantly involves a non-trivial central extension.  {A concluding section (Section {\bf \ref{Sec:Conclusions}}) discusses the generalization to higher dimensions and points out which supertranslations at spatial infinity have been discussed  at null infinity. }Two appendices closes our paper.  In \ref{App:Comparison} we briefly compare our approach to asymptotic symmetries with other approaches carried also at spatial infinity.  In \ref{App:Poinc}, we indicate how Poincar\'e invariance (with our set of boundary conditions) is proved and discuss the algebra of the asymptotic symmetries. 

\section{Linearized gravity in $6$ spacetime dimensions}
\label{Sec:Presentation}

\subsection{Action and gauge symmetries}

The Hamiltonian action of linearized gravity on Minkowski spacetime
read
\begin{equation}
I=\int dt\left[\int d^{5}x\left(\pi^{ij}\dot{h}_{ij}-\mathcal{E}-n\mathcal{G}-n^{i}\mathcal{G}_{i}\right) - B_{\infty}\right]\,,\label{eq:Action}
\end{equation}
Here, the dynamical fields are the canonically conjugate pairs $(\pi^{ij},h_{ij})$,
while $n$ and $n^{i}$ are the Lagrange multipliers associated
with the following constraints
\begin{eqnarray}
\mathcal{G} && =\sqrt{g}(\triangle h-\nabla^{i}\nabla^{j}h_{ij})\,,\label{eq:HamG}\\
\mathcal{G}_{i} && =-2\nabla^{j}\pi_{ij}\,.\label{eq:MomG}
\end{eqnarray}
The covariant derivatives are the flat space ones, i.e., $\nabla_i = \partial_i$ in cartesian coordinates.  The term $B_{\infty}$ is a surface term at infinity that might be needed to supplement the bulk time generator $\int d^{5}x \left(\mathcal{E} + n\mathcal{G} + n^{i}\mathcal{G}_{i}\right)$.  Its explicit form depends on the asymptotic behaviour of  the Lagrange multipliers and will be discussed below. 

The constraints are first class and generate the infinitesimal gauge transformations
\begin{eqnarray}
\delta_{\epsilon}h_{ij} && =\nabla_{i}\epsilon_{j}+\nabla_{j}\epsilon_{i}\,,\label{eq:dh-Gauge}\\
\delta_{\epsilon}\pi^{ij} && =\sqrt{g}\left(\nabla^{i}\nabla^{j}\epsilon-g^{ij}\triangle\epsilon\right)\,.\label{eq:dp-Gauge}
\end{eqnarray}
under which the Hamiltonian action (\ref{eq:Action}) is invariant.  
The gauge transformations (\ref{eq:dh-Gauge})-(\ref{eq:dp-Gauge}) are just the standard diffeomorphisms linearized around the Minkowski background.  For that reason, they will also be called ``(linearized) diffeomorphisms''.  They are obtained by taking the Poisson brackets of the dynamical variables with the integral $\int d^{5}x \left(\epsilon \mathcal{G} + \epsilon^{i}\mathcal{G}_{i}\right) + Q[\epsilon, \epsilon^i]$, where $Q[\epsilon, \epsilon^i]$ is a surface term at infinity which will also be given below, but which does not contribute to the local variations  (\ref{eq:dh-Gauge})-(\ref{eq:dp-Gauge}) of the canonical variables. 

Because (\ref{eq:dh-Gauge})-(\ref{eq:dp-Gauge}) are the diffeomorphisms linearized around the Minkowski background, they vanish identically for the Minkowski isometries, in particular when $\epsilon_{i} = $ constant and $\epsilon = $ constant.

The energy and momentum densities of the theory are given by 
\begin{eqnarray}
\mathcal{E} & = & \frac{1}{\sqrt{g}}\left(\pi^{ij}\pi_{ij}-\frac{\pi^{2}}{4}\right)+\sqrt{g}\left(\frac{1}{4}\nabla_{k}h_{ij}\nabla^{k}h^{ij}-\frac{1}{2}\nabla_{j}h^{ij}\nabla^{k}h_{ik}+\frac{1}{4}\nabla_{i}h\nabla^{i}h\right) \nonumber \\
 &  & +\sqrt{g}\nabla_{l}\left(-h^{ij}\nabla^{l}h_{ij}-h^{il}\nabla_{i}h+\frac{3}{2}h^{lj}\nabla^{i}h_{ij}+\frac{1}{2}h_{ij}\nabla^{i}h^{jl}\right)+\frac{1}{2}h\mathcal{G}\,,\\
\mathcal{P}_{i} & = & -2\partial_{j}\left(\pi^{jk}h_{ik}\right)+\pi^{jk}\partial_{i}h_{jk}\,,
\end{eqnarray}
respectively. \newline  \newline 

The  action (\ref{eq:Action}) is also invariant
under Poincar\'e transformations (see \ref{App:Poinc}), which takes  the Hamiltonian form
\begin{eqnarray}
\delta_{\xi}h_{ij} & = & \frac{2\xi}{\sqrt{g}}\left(\pi_{ij}-\frac{1}{4}g_{ij}\pi\right)+\mathcal{L}_{\xi}h_{ij}\,,\label{eq:dh-Poincare}\\
\delta_{\xi}\pi^{ij} & = & \frac{1}{2}\sqrt{g}\xi\left(\triangle h^{ij}+\nabla^{i}\nabla^{j}h-2\nabla^{(i}\nabla_{k}h^{j)k}\right) \nonumber \\
& & +\frac{1}{2}\sqrt{g}\nabla_{k}\xi\left[\nabla^{k}h^{ij}-2\nabla^{(i}h^{j)k}+g^{ij}\left(2\nabla_{l}h^{kl}-\nabla^{k}h\right)\right]\nonumber \\
 &  & +\sqrt{g}\triangle\xi h^{ij}+\sqrt{g}g^{ij}\nabla_{k}\nabla_{l}\xi h^{kl}-2\sqrt{g}\nabla_{k}\nabla^{(i}\xi h^{j)k} \nonumber \\
 & & -\frac{1}{2}\sqrt{g}\left(\nabla^{i}\nabla^{j}\xi-g^{ij}\triangle\xi\right)h-\frac{1}{2}g^{ij}\xi\mathcal{G}+\mathcal{L}_{\xi}\pi^{ij}\,,\label{eq:dp-Poincare}
\end{eqnarray}
where the spatial Lie derivatives of $h_{ij}$ and $\pi^{ij}$ read
respectively as
\begin{eqnarray}
\mathcal{L}_{\xi}h_{ij} && =2h_{k(i}\partial_{j)}\xi^{k}+\xi^{k}\partial_{k}h_{ij}\,,\\
\mathcal{L}_{\xi}\pi^{ij} && =-2\partial_{k}\xi^{(i}\pi^{j)k}+\partial_{k}(\xi^{k}\pi^{ij})\,.
\end{eqnarray}
Here $\xi$ and $\xi^i$  are the normal and tangential components of the vector fields defining the Poincar\'e transformations,
\begin{equation}
\xi = b_i x^i + a^0 \, , \qquad \xi^i = {b^i}_j x^j + a^j \, , \qquad b_{ij} = - b_{ji} \, .
\end{equation}

\subsection{Asymptotic conditions - Preliminary form\label{Asymptotia0}}

The natural fall-off of the dynamical fields in $5+ 1$ dimensions is, in asymptotically cartesian coordinates, $h_{ij} \sim r^{-3}$, $\pi^{ij} \sim r^{-4}$.  These conditions are fulfilled by the Myers-Perry solutions \cite{Myers:1986un}. It might therefore be tempting to impose these conditions at infinity,
namely,
$$h_{ij} = \mathcal O(r^{-3}) \, , \quad \pi^{ij} = \mathcal O(r^{-4})$$
for $r \rightarrow \infty$. 
 However, these conditions, being expressed in terms of non-diffeomorphism invariant fields, also implicitly fix the coordinate system at infinity. This gauge fixing might be ``improper'', i.e., might gauge away a freedom with physical content, which does not corrrespond to a mere redundancy in the description of the system.  This was found to be the case in $3+1$ \cite{Henneaux:2018hdj,Henneaux:2019yax} and $4+1$ \cite{Fuentealba:2021yvo,Fuentealba:2022yqt} dimensions. 

More precisely, the boundary conditions $h_{ij} = \mathcal O(r^{-3})$, $\pi^{ij} = \mathcal O(r^{-4})$ leave as only coordinate freedom $\mathcal O(r^{-2})$ gauge transformations.  In particular angle-dependent $\mathcal O(1)$ gauge transformations are not permitted.  But these precisely correspond to supertranslations.

We should therefore formulate the boundary conditions in a way that does not freeze the possibility to perform non-trivial $\mathcal O(1)$ gauge transformations.  The simplest way to implement this requirement is to act with such gauge transformations on solutions with the behaviour $h_{ij} = \mathcal O(r^{-3})$, $\pi^{ij} = \mathcal O(r^{-4})$,  and analyse what new terms are generated.  The boundary conditions should allow these new terms.

Now, it was found both in $3+1$ dimensions \cite{Henneaux:2018hdj,Henneaux:2019yax} and in $4+1$ dimensions \cite{Fuentealba:2021yvo,Fuentealba:2022yqt} that not all $\mathcal O(1)$ gauge transformations could be incorporated into the formalism, but only a subclass of them - sufficient to contain the standard supertranslations in $3+1$ dimensions. Non trivial constraints arise from Lorentz invariance, which might be violated by surface terms, as well as from the finiteness of the symplectic form and of the charges, which are not expressed in terms of gauge invariant curvatures only and so ``see'' the improper gauge transformations. 

There might be many ways to consistently restrict the class of allowed diffeomorphisms at infinity, leading to different asymptotic descriptions.  Indeed, in $3+1$ dimensions, there are the boundary conditions of \cite{Henneaux:2018hdj,Henneaux:2019yax}, the extended boundary conditions of \cite{Fuentealba:2022xsz} allowing logarithmic supertranslations, and the inequivalent boundary conditions of \cite{Henneaux:2018cst} leading to a null infinity with more serious logarithmic singularities (although perfectly fine on Cauchy hypersurfaces).  And one might perhaps suspect that there are even more possibilities.

We shall consider here general gauge transformations of order $\mathcal O(1)$, without a priori  restrictions, and exhibit as we proceed  the extra conditions that must be imposed for finiteness.
We thus adopt tentatively the following boundary conditions on $h_{ij}$ (in polar coordinates)
\begin{eqnarray}
h_{rr} && =\frac{1}{r^{2}}\lambda^{(0)}+\frac{1}{r^{3}}\xbar h_{rr}+\frac{1}{r^{4}}h_{rr}^{(2)}+o(r^{-4})\,, \nonumber \\ 
h_{rA} && =\lambda_{A}^{(0)} + \frac{1}{r}\lambda_{A}^{(1)}+\frac{1}{r^{2}}\xbar h_{rA}+\frac{1}{r^{3}}h_{rA}^{(2)}+o(r^{-3})\,, \nonumber \\ 
h_{AB} && =r\theta_{AB}^{(0)}+\theta_{AB}^{(1)}+\frac{1}{r}\xbar h_{AB}+\frac{1}{r^{2}}h_{AB}^{(2)}+o(r^{-2})\,, \nonumber  
\end{eqnarray}
while the asymptotic conditions of its conjugate momentum $\pi^{ij}$ are taken to be (recalling that $\pi^{ij}$ carries density weight one),
\begin{eqnarray}
\pi^{rr} && =r^{2}\kappa_{(0)}^{rr}+r\kappa_{(1)}^{rr}+\xbar\pi^{rr}+\frac{1}{r}\pi_{(2)}^{rr}+o(r^{-1})\,,\nonumber \\ 
\pi^{rA} && =r\kappa_{(0)}^{rA}+\kappa_{(1)}^{rA}+\frac{1}{r}\xbar\pi^{rA}+\frac{1}{r^{2}}\pi_{(2)}^{rA}+o(r^{-2})\,,\nonumber \\ 
\pi^{AB} && =\kappa_{(0)}^{AB}+\frac{1}{r}\kappa_{(1)}^{AB}+\frac{1}{r^{2}}\xbar\pi^{AB}+\frac{1}{r^{3}}\pi_{(2)}^{AB}+o(r^{-3})\,. \nonumber 
\end{eqnarray}
Here, we have set
\begin{eqnarray}
\lambda^{(0)} && =-2F^{(0)}\,, \nonumber \\
\lambda_{A}^{(0)} && =\frac{1}{2}\xbar D_{A}U-G_{A}\,,\nonumber \\
\lambda_{A}^{(1)} && =\xbar D_{A}F^{(0)}-2M_{A}^{(0)}\,, \nonumber \\
\theta_{AB}^{(0)} && =\xbar D_{A}G_{B}+\xbar D_{B}G_{A}+\xbar g_{AB}U\,, \nonumber \\
\theta_{AB}^{(1)} && =\xbar D_{A}M_{B}^{(0)}+\xbar D_{B}M_{A}^{(0)}+2\xbar g_{AB}F^{(0)}\,, \nonumber
\end{eqnarray}
\begin{eqnarray}
\kappa_{(0)}^{rr} && =\sqrt{\xbar g}\xbar\triangle V\,, \nonumber \\
\kappa_{(1)}^{rr}& & =\sqrt{\xbar g}\left(\xbar\triangle N^{(0)}-4N^{(0)}\right)\,, \nonumber
\end{eqnarray}
\begin{eqnarray}
\kappa_{(0)}^{rA} && =\sqrt{\xbar g}\xbar D^{A}V\,, \nonumber \\
\kappa_{(1)}^{rA} && =2\sqrt{\xbar g}\xbar D^{A}N^{(0)}\,, \nonumber
\end{eqnarray}
\begin{eqnarray}
\kappa_{(0)}^{AB} && =\sqrt{\xbar g}\left(\xbar g^{AB}\xbar\triangle V-\xbar D^{A}\xbar D^{B}V\right)\,, \nonumber \\
\kappa_{(1)}^{AB} && =\sqrt{\xbar g}\left(\xbar g^{AB}\xbar\triangle N^{(0)}-\xbar g^{AB}N^{(0)}-\xbar D^{A}\xbar D^{B}N^{(0)}\right)\,, \nonumber
\end{eqnarray}
where the functions $F^{(0)}$, $G_A$, $M_{A}^{(0)}$, $U$, $V$ and $N^{(0)}$ are at this stage arbitrary functions
on the 4-sphere at spatial infinity.

The terms with an overbar, such as $\frac{1}{r^{3}}\xbar h_{rr}$ in $h_{rr}$ or $\xbar\pi^{rr}$ in $\pi^{rr}$ are the terms corresponding to $h_{ij} = \mathcal O(r^{-3})$, $\pi^{ij} = \mathcal O(r^{-4})$ when expressed in polar coordinates.  These terms are not subject to the parity conditions present in $3+1$ dimensions, where they are the leading terms.   Indeed, as shown below, they do not lead to divergences in the symplectic structure in $5+1$ dimensions even in the absence of definite parity conditions, which should thus  not be assumed.   In particular,  their quadratic contribution to the kinetic term, which is the dangerous one in $3+1$ dimensions,  is now $ \sim r^4 dr r^{-4} r^{-3} \sim dr r^{-3}$ and just converges as such, as it happens already in $4+1$ dimensions (no logarithmic divergence that must be cancelled).

The terms occuring at a lower power of $r^{-1}$ in the asymptotic expansion are generated by gauge transformations (\ref{eq:dh-Gauge})-(\ref{eq:dp-Gauge}) of order $\mathcal O (1)$. 
They take  the form
\begin{eqnarray}
\epsilon && = - V -\frac{1}{r}N^{(0)}+ \mathcal O(r^{-2})\,,\\
\epsilon^{r} && =\frac12 U+\frac{1}{r}F^{(0)} + \mathcal O(r^{-2})\,,\\
\epsilon^{A} && =\frac{1}{r} G_A +\frac{1}{r^{2}}M_{(0)}^{A}+ \mathcal O(r^{-3})\,.
\end{eqnarray}
The minus signs in the first line and the factor of $\frac12$ in the second line have no particular significance and are chosen for the sole purpose of simplifying some of the subsequent formulas. 

We note that the zero mode of $V$ and the $l=1$-component of $U$ in an expansion in terms of spherical harmonics drop out from the asymptotic expansion of the metric components and their conjugate momenta. This is as it should since they correspond to $\epsilon_{i} = $ constant and $\epsilon = $ constant.  A useful check below will be that the charges of the gauge transformations vanish for such choices of $V$ and $U$.  

As we shall now show, these boundary conditions will have to be strengthened for consistency.  The final form of the boundary conditions is given in Section {\bf \ref{Asymptotia1}}.

 \subsection{Asymptotic form of the constraints\label{constraints}}

The asymptotic expansion of the constraint (\ref{eq:HamG})
 reads
\begin{equation}
\mathcal{G}=r\mathcal{G}^{(-1)}+\mathcal{G}^{(0)}+\frac{1}{r}\mathcal{G}^{(1)}+o(r^{-1})\,,
\end{equation}
where
\begin{align}
\mathcal{G}^{(-1)} & = \sqrt{\xbar g}\left(\xbar\triangle\theta^{(0)}-\xbar D_{A}\xbar D_{B}\theta^{(0)AB}\right),\\
\mathcal{G}^{(0)} & = \sqrt{\xbar g}\left(\xbar\triangle\theta^{(1)}-\xbar D_{A}\xbar D_{B}\theta^{(1)AB}-\theta^{(1)}-4\xbar D_{A}\lambda^{(1)A}+\xbar\triangle\lambda^{(0)}-4\lambda^{(0)}\right),\\
\mathcal{G}^{(1)} & = \sqrt{\xbar g}\left(\xbar\triangle\,\xbar h+\xbar\triangle\,\xbar h_{rr}-\xbar D_{A}\xbar D_{B}\xbar h^{AB}-2\xbar D_{A}\xbar h_{r}^{A}\right).
\end{align}
$\mathcal{G}^{(-1)}$ and $\mathcal{G}^{(0)}$ are identically zero
because of the definitions of $\theta_{AB}^{(0)}$, $\theta_{AB}^{(1)}$,
$\lambda_{A}^{(1)}$ and $\lambda^{(0)}$.

The  ``momentum'' constraint (\ref{eq:MomG}) goes asymptotically as
\begin{eqnarray}
\mathcal{G}_{r} && =2r\left(-2\kappa_{(0)}^{rr}+\kappa_{(0)}-\xbar D_{A}\kappa_{(0)}^{rA}\right)-2\left(\kappa_{(1)}^{rr}-\kappa_{(1)}+\xbar D_{A}\kappa_{(1)}^{rA}\right) \nonumber \\
&& \quad +\frac{2}{r}\left(\xbar\pi-\xbar D_{A}\xbar\pi^{rA}\right)+\mathcal{O}\left(r^{-2}\right)\,,\\
\mathcal{G}_{A} && =-2r^{2}\left(3\kappa_{(0)A}^{r}+\xbar D_{B}\kappa_{(0)A}^{B}\right)-2r\left(2\kappa_{(1)A}^{r}+\xbar D_{B}\kappa_{(1)A}^{B}\right) \nonumber \\
&& \quad -2\left(\xbar\pi_{A}^{r}+\xbar D_{B}\xbar\pi_{A}^{B}\right)+\mathcal{O}\left(r^{-1}\right)\,.
\end{eqnarray}
The leading and subleading terms of $\mathcal{G}_{r}$ vanish identically on account of the special form of the first terms in the expansion of the fields.
The same happens with the leading and subleading terms of $\mathcal{G}_{A}.$

\section{Finiteness of the symplectic structure \label{FinitenessSymplectic}}

Replacing the asymptotic expansions of the canonical variables into the kinetic term, we get
\begin{eqnarray}
&& K  =\int dt \int dr \oint d^{4}x\Big[r \left(\kappa_{(0)}^{AB}\dot{\theta}_{AB}^{(0)}+2\kappa_{(0)}^{rA}\dot{\lambda}_{A}^{(0)}\right) \nonumber \\
 && \quad\qquad\,\,+\kappa_{(0)}^{rr}\dot{\lambda}^{(0)}+2\kappa_{(0)}^{rA}\dot{\lambda}_{A}^{(1)}+\kappa_{(0)}^{AB}\dot{\theta}_{AB}^{(1)}+\kappa_{(1)}^{AB}\dot{\theta}_{AB}^{(0)} +2\kappa_{(1)}^{rA}\dot{\lambda}_{A}^{(0)} \nonumber \\
 && \quad\qquad\,\,+\frac{1}{r}\Big(\kappa_{(1)}^{rr}\dot{\lambda}^{(0)}+2\kappa_{(1)}^{rA}\dot{\lambda}_{A}^{(1)}+\kappa_{(1)}^{AB}\dot{\theta}_{AB}^{(1)}+\kappa_{(0)}^{rr}\dot{\xbar h}_{rr} \nonumber \\
 && \quad\qquad\,\,+2\kappa_{(0)}^{rA}\dot{\xbar h}_{rA}+\kappa_{(0)}^{AB}\dot{\xbar h}_{AB}+\xbar\pi^{AB}\dot{\theta}_{AB}^{(0)}+2\xbar\pi^{rA}\dot{\lambda}_{A}^{(0)}\Big)+\mathcal{O}\left(r^{-2}\right)\Big]\, ,
\end{eqnarray}
an expression that can be used to investigate the behaviour of the integral $\int^R dr$ as $r \rightarrow \infty$.   

Using the definition of $\kappa_{(0)}^{ij}$, the coefficient of the quadratic divergence ($\int^R dr \,  r \sim R^2 $) can  be written, after an integration by parts, as
\begin{equation}
\oint d^{4}x\sqrt{\xbar g}\,V\left(-2\xbar D_{A}\dot{\lambda}^{(0)A}+\xbar\triangle\theta^{(0)}-\xbar D_{A}\xbar D_{B}\theta^{(0)AB}\right)\,,
\end{equation}
The identity $\mathcal{G}^{(1)}=0$ can then be used to reduce this expression to
\begin{equation}
4\oint d^{4}x\sqrt{\xbar g}\,V\xbar D_{A}\dot{\lambda}^{(0)A}\,,
\end{equation}
which vanishes provided one imposes the condition
\begin{equation}
\xbar D_{A}\lambda^{(0)A}=0\quad\Leftrightarrow\quad\xbar D_{A}G^{A}=\frac{1}{2}\xbar\triangle\,U\,.
\end{equation}

For the linear divergence, we make use of the definitions of $\kappa_{(0)}^{ij}$
and  $\kappa_{(1)}^{ij}$. After integration by parts,
the associated term becomes
\begin{eqnarray}
&&\oint d^{4}x\sqrt{\xbar g}\Big[V\left(\xbar\triangle\dot{\theta}^{(1)}-\xbar D_{A}\xbar D_{B}\dot{\theta}^{(1)AB}-2\xbar D_{A}\dot{\lambda}^{(1)A}+\xbar\triangle\dot{\lambda}^{(0)}\right) \nonumber \\
&&+N^{(0)}\left(\xbar\triangle\dot{\theta}^{(0)}-\xbar D_{A}\xbar D_{B}\dot{\theta}^{(0)AB}-2\xbar D_{A}\dot{\lambda}^{(0)A} -\dot{\theta}^{(0)}\right)\Big]\,.
\end{eqnarray}
We now make use of the identities $\mathcal{G}^{(1)}=0=\mathcal{G}^{(0)}$ and 
find that the above integral reduces to
\begin{equation}
\oint d^{4}x\sqrt{\xbar g}\left[V\left(\dot{\theta}^{(1)}+2\xbar D_{A}\dot{\lambda}^{(1)A}+4\dot{\lambda}^{(0)}\right)-N^{(0)}\dot{\theta}^{(0)}\right]\,.
\end{equation}
Using the definition of the functions, we get, after integration
by parts in the second term, the final expression
\begin{equation}
\oint d^{4}x\sqrt{\xbar g}\left[2V\left(\xbar\triangle\,\dot{F}^{(0)}-\xbar D_{A}\dot{M}^{(0)A}\right)-\dot{U}\left(\xbar\triangle N^{(0)}+4N^{(0)}\right)\right]\,.
\end{equation}
Since $U$ and $V$ are required to be arbitrary,  this integral is zero provided
\begin{equation}
\xbar\triangle\,F^{(0)}=\xbar D_{A}M^{(0)A}\quad,\qquad\left(\xbar\triangle+4\right)N^{(0)}=0\,.
\end{equation}
 Note that the second condition expresses that  $N^{(0)}$ can only have components along the $l=1$ spherical harmonics.

Finally, for the logarithmic divergence we make use of the definitions
of $\kappa_{(0)}^{ij}$, $\kappa_{(1)}^{ij}$, $\theta_{AB}^{(0)}$
and $\lambda_{A}^{(0)}$ to obtain the following integral
\begin{eqnarray}
&&\oint d^{4}x\sqrt{\xbar g}\Big[N^{(0)}\left(\xbar\triangle\dot{\theta}^{(1)}-\xbar D_{A}\xbar D_{B}\dot{\theta}^{(1)AB}-\dot{\theta}^{(1)}-4\xbar D_{A}\dot{\lambda}^{(1)A}+\xbar\triangle\dot{\lambda}^{(0)}-4\dot{\lambda}^{(0)}\right)\nonumber \\
&&+V\left(\xbar\triangle\,\dot{\xbar h}+\xbar\triangle\,\dot{\xbar h}_{rr}-\xbar D_{A}\xbar D_{B}\dot{\xbar h}^{AB}-2\xbar D_{A}\dot{\xbar h}_{r}^{A}\right) \nonumber \\
&&-\left(\xbar\pi^{rA}+\xbar D_{B}\xbar\pi^{AB}\right)D_{A}\dot{U}+\left(\xbar\pi-\xbar D_{A}\xbar\pi^{rA}\right)\dot{U}\Big]\,.
\end{eqnarray}
The first line vanishes because of the identity $\mathcal{G}^{(0)}=0$,
while the second and third lines are zero if we impose, as in  \cite{Henneaux:2018hdj}, the faster fall-off
of the constraints
\begin{equation}
\mathcal{G}=\mathcal{O}\left(r^{-2}\right)\,,\quad\mathcal{G}_{r}=\mathcal{O}\left(r^{-2}\right)\,,\quad\mathcal{G}_{A}=\mathcal{O}\left(r^{-1}\right)\,.
\end{equation}
This completes the proof that the kinetic term is finite.

\section{Canonical generator of the gauge symmetries}
\label{Sec:FinitessGen}

\subsection{Conditions for finiteness}

We now turn to the question of the finiteness of the canonical generators of the asymptotic symmetries, which are given by
\begin{equation}
G[\epsilon,\epsilon^{i}] =\int d^{5}x\left(\epsilon\mathcal{G}+\epsilon^{i}\mathcal{G}_{i}\right)+Q[\epsilon,\epsilon^{i}]\,,
\end{equation}
where we adopt a general $\mathcal O(1)$ behaviour for the gauge parameters,
\begin{eqnarray}
\epsilon && =T+\frac{1}{r}T^{(1)}+\frac{1}{r^{2}}T^{(2)}+o(r^{-2})\,,\\
\epsilon^{r} && =W+\frac{1}{r}W^{(1)}+\frac{1}{r^{2}}W^{(2)}+o(r^{-2})\,,\\
\epsilon^{A} && =\frac{1}{r}I^{A}+\frac{1}{r^{2}}I_{(1)}^{A}+\frac{1}{r^{3}}I_{(2)}^{A}+o(r^{-3})\,,
\end{eqnarray}
with $T$, $T^{(1)}$, $T^{(2)}$, $W$, $W^{(1)}$, $W^{(2)}$, $I^{A}$, $I_{(1)}^{A}$ and $I_{(2)}^{A}$ at this stage arbitrary functions on the sphere.  Recall that $\mathcal O(1)$ terms in $\epsilon^i$ appear at order $\mathcal O(r^{-1})$ in the angular components $\epsilon^A$. 

We will find that finiteness of $G[\epsilon,\epsilon^{i}]$ imposes the same restrictions on these functions as the ones found in the previous section.  We have explicitly written the $T^{(2)}$, $W^{(2)}$  and $I_{(2)}^{A}$ terms because these can define improper gauge symmetries with non-vanishing charges, even though they affect only the subleading terms  in the asymptotic expansion of the canonical fields (those with an overbar).

The surface integral $Q[\epsilon,\epsilon^{i}]$ that accompanies the weakly vanishing bulk term $\int d^{5}x\left(\epsilon\mathcal{G}+\epsilon^{i}\mathcal{G}_{i}\right)$ is determined through the central equation $\iota_X \Omega = - d_V G$ (see \cite{Henneaux:2018gfi} for notations and more information), which reduces in the present case to the criterion of \cite{Regge:1974zd}, namely, that $G[\epsilon,\epsilon^{i}]$ should have well-defined functional derivatives.
This yields the condition
\begin{equation}
\delta Q[\epsilon,\epsilon^{i}]=\oint d^{4}s_{i}\Big[\sqrt{g}\epsilon\nabla_{j}\left(\delta h^{ij}-g^{ij}\delta h\right)-\sqrt{g}\nabla_{j}\epsilon\left(\delta h^{ij}-g^{ij}\delta h\right)+2\epsilon_{j}\delta\pi^{ij}\Big] , \label{eq:dQG-1}
\end{equation}
where $\delta h_{ij}$ and $\delta \pi^{ij}$ are the variations of the fields under the gauge transformations (\ref{eq:dh-Gauge})-(\ref{eq:dp-Gauge}).

Replacing the explicit form of the asymptotic expansions of the fields in the above surface integral for $\delta Q[\epsilon,\epsilon^{i}] \equiv \delta Q$, 
we get the expression
\begin{eqnarray}
\delta Q && =r^{2}\oint d^{4}x\left(-2\sqrt{\xbar g}T\xbar D_{A}\delta\lambda^{(0)A}+2W\delta\kappa_{(0)}^{rr}+2I_{A}\delta\kappa_{(0)}^{rA}\right) \nonumber\\
 && \quad+r\oint d^{4}x\Big[\sqrt{\xbar g}T\left(4\delta\lambda^{(0)}+\delta\theta^{(1)}+2\xbar D_{A}\delta\lambda^{(1)A}\right)+\sqrt{\xbar g}T^{(1)}\left(-\delta\theta^{(0)}+2\xbar D_{A}\delta\lambda^{(0)A}\right) \nonumber\\
 && \quad\qquad\qquad\qquad+2W\delta\kappa_{(1)}^{rr}+2I_{A}\delta\kappa_{(1)}^{rA}+2W^{(1)}\delta\kappa_{(0)}^{rr}+2I_{A}^{(1)}\delta\kappa_{(0)}^{rA}\Big] \nonumber\\
 && \quad+\oint d^{4}x\Big[\sqrt{\xbar g}T\left(4\delta\xbar h_{rr}+2\delta\xbar h+2\xbar D_{A}\delta\xbar h_{r}^{A}\right)+\sqrt{\xbar g}T^{(1)}\left(4\delta\lambda^{(0)}+2\xbar D_{A}\delta\lambda^{(1)A}\right) \nonumber\\
 && \quad\qquad\qquad\qquad+\sqrt{\xbar g}T^{(2)}\left(-2\delta\theta^{(0)}+2\xbar D_{A}\delta\lambda^{(0)A}\right)+2W\delta\xbar\pi{}^{rr}+2I_{A}\delta\xbar\pi^{rA} \nonumber\\
 && \quad\qquad\qquad\qquad+2W^{(1)}\delta\kappa_{(1)}^{rr}+2I_{A}^{(1)}\delta\kappa_{(1)}^{rA}+2W^{(2)}\delta\kappa_{(0)}^{rr}+2I_{A}^{(2)}\delta\kappa_{(0)}^{rA}\Big]\,. \nonumber
\end{eqnarray}

In order to establish finiteness of the variation of the charge we examine each contribution in turn:
\begin{itemize}
\item In the quadratic divergence, the term proportional to $T$ vanishes
if we impose the condition $\xbar D_{A}\lambda^{(0)A}=0$, which is the same condition found to be necessary for eliminating the quadratic divergence in the kinetic term. Consistency with gauge transformations
$\delta_{\epsilon}\lambda_{A}^{(0)}=\xbar D_{A}W-I_{A}$ implies
that the gauge parameters should be restricted as $\xbar D_{A}I^{A}=\xbar\triangle W$. This latter relation kills
the second and third terms after using that $\kappa_{(0)}^{rr}=\sqrt{\xbar g}\xbar\triangle V$
and $\kappa_{(0)}^{rA}=\sqrt{\xbar g}\xbar D^{A}V$, as well as integration
by parts.
\item {The stronger condition $\lambda^{(0)A}= 0$, which clearly implies $\xbar D_{A}\lambda^{(0)A}=0$,   turns out to be needed in the proof of Lorentz invariance  (see \ref{App:Poinc}).   We shall thus impose from now that condition,
\be
\lambda^{(0)}_A = 0 \quad \Leftrightarrow \quad G_A = \frac12 \xbar D_A U \, .
\ee
This restricts the gauge transformations to
\be
I_A = \xbar D_A W \, .
\ee}
\item In the linear divergence, the term proportional to 
$T$ vanishes if we impose the condition $\xbar\triangle\,F^{(0)}=\xbar D_{A}M^{(0)A}$, again found necessary above for finiteness of the symplectic form.
Consistency of the latter condition with gauge transformations implies that the gauge parameters should be such that $\xbar D_{A}I^{(1)A}=\xbar\triangle\,W^{(1)}$.
This restriction automatically kills the terms proportional to $I^{(1)A}$ and  $\xbar\triangle\,W^{(1)}$ (which are also those containing $\delta V$) as can be seen through an integration by parts on the sphere.
Similarly, the term proportional to $W$ vanishes by imposing the condition
$\left(\xbar\triangle+4\right)N^{(0)}=0$ found above. Again consistency of the
latter condition with the gauge transformations implies that $\left(\xbar\triangle+4\right)T^{(1)}=0$,
which is the condition needed to make zero the remaining term proportional to $T^{(1)}$ (using the condition  $\xbar D_{A}\lambda^{(0)A}=0$, the definition of $\theta^{(0)}$ and  integrations
by parts).
\end{itemize}
Thus, all the divergent terms are actually zero and the variation of the charge is finite under the same conditions that ensure finiteness of the kinetic term.

Integrability is automatic for field-independent parameters (as here) since $\delta Q$ is linear.  One finds that $Q$ 
is given by
\begin{eqnarray}
Q && =\oint d^{4}x\Big[\sqrt{\xbar g}T\left(4\xbar h_{rr}+2\xbar h+2\xbar D_{A}\xbar h_{r}^{A}\right)-2\sqrt{\xbar g}T^{(2)}\theta^{(0)} \nonumber \\
&& \qquad \qquad +2W\xbar\pi^{rr}+2I_{A}\xbar\pi^{rA}+2W^{(2)}\kappa_{(0)}^{rr}+2I_{A}^{(2)}\kappa_{(0)}^{rA}\Big]\,.
\end{eqnarray}
This expression can can be re-written as
\begin{eqnarray}
Q && =\oint d^{4}x\Big[\sqrt{\xbar g}T\left(4\xbar h_{rr}+2\xbar h+2\xbar D_{A}\xbar h_{r}^{A}\right)+2W\left(\xbar\pi^{rr}-\xbar D_{A}\xbar\pi^{rA}\right) \nonumber \\
&& \qquad \qquad -2\sqrt{\xbar g}\tilde{T}^{(2)}U-2\tilde{W}^{(2)}V\Big]\,, \label{Eq:ExprForQ}
\end{eqnarray}
where
\begin{eqnarray}
\tilde{T}^{(2)} && =\left(\xbar\triangle+4\right)T^{(2)}\,,  \\
\tilde{W}^{(2)} && =\xbar D_{A}I^{(2)A}-\xbar\triangle W^{(2)}\, .
\end{eqnarray}

\subsection{Discussion}

Various important facts are manifest from (\ref{Eq:ExprForQ}):
\begin{itemize}
\item The zero mode of $T$ and the $l=1$ mode of $W$ both give a zero charge.  Indeed, as we have observed, the first terms in the asymptotic expansion of the fields drop from the constraints.  If we denote by $h'_{ij}$ and $\pi'^{ij}$ the fields obained by truncating the asymptotic expansions up to the terms with an overbar (which are kept), one finds
\begin{eqnarray}
\mathcal{G} && =\sqrt{g}(\triangle h-\nabla^{i}\nabla^{j}h_{ij})    =\sqrt{g}(\triangle h'-\nabla^{i}\nabla^{j}h'_{ij})  \nonumber\\
\mathcal{G}_{i} && =-2\nabla^{j}\pi_{ij} =  -2\nabla^{j}\pi'_{ij} \, . \nonumber
\end{eqnarray}  
Consequently, if we take $\epsilon = T =$ constant $\equiv a$ and $\epsilon^i = $ constant $\equiv a^i$, implying $W = a^i Y_i^1$, we get
\begin{eqnarray}
G[\epsilon,\epsilon^{i}] && =\int d^{5}x\Big[a\sqrt{g}(\triangle h'-\nabla^{i}\nabla^{j}h'_{ij})+a^{i}( -2\nabla^{j}\pi'_{ij} )\Big]\nonumber \\
&& \quad +\oint d^{4}x\Big[\sqrt{\xbar g}a\left(4\xbar h_{rr}+2\xbar h+2\xbar D_{A}\xbar h_{r}^{A}\right)+2a^i Y_i^1\xbar\pi^{rr}\Big]\, ,   \nonumber
\end{eqnarray}
an expression that can be seen to identically vanish by converting the surface integral into a bulk integral through the use of Stokes theorem. 
\item No $\mathcal O(r^{-1})$ coefficients $T^{(1)}$,  $W^{(1)}$ or $I_{(1)}^{A}$ of the gauge transformations appear in the charges (\ref{Eq:ExprForQ}).  This means that these gauge transformations are proper gauge transformations.  They can be eliminated by imposing the proper gauge conditions $N^{(0)} = F^{(0)} = M^A_{(0)} = 0$.   The only gauge transformations that preserve these conditions have $T^{(1)} = W^{(1)}= I_{(1)}^{A} = 0$. {A weaker gauge condition is 
\be
M_{(0)}^{A} = \xbar D^A F^{(0)}
\ee
preserved by gauge transformations such that $ I_{(1)}^{A} = \xbar D^A W^{(1)}$.}

\item Finally, the $\mathcal O(r^{-2})$ coefficients do contribute to the charges and define accordingly improper gauge symmetries. They correspond to the $D=4$ logarithmic gauge transformations of \cite{Fuentealba:2022xsz}. Their charges are the functions $U$ and $V$, that parametrize the $\mathcal O(1)$ gauge transformation terms in the dynamical fields -- terms that are related to the memory effect in the Hamiltonian formulation \cite{Fuentealba:2023syb}.  Because of the projection operators appearing in $\tilde{T}^{(2)}$ and $\tilde{W}^{(2)}$, neither the zero mode of $V$ nor the $l=1$ mode of $U$ contribute to the charge.  Furthermore, since only the combination $\tilde{W}^{(2)}$ of the gauge parameters $I^{(2)A}$ and $ W^{(2)}$ appears in the charge,  $\mathcal O(r^{-2})$ spatial gauge transformations that have zero  $\tilde{W}^{(2)}$ are proper gauge transformations. 
\end{itemize}
 
 The conclusion of our analysis is that only the $\mathcal O(1)$ and $\mathcal O(r^{-2})$ terms in the asymptotic expansion of the gauge parameters (with the restrictions on the spherical harmonic components and on the angular components described above) define improper gauge transformations having a non-trivial action on the system.  The $\mathcal O(r^{-1})$ terms correspond to proper gauge symmetries and can be gauged away.  The improper gauge transformations are described by four independent functions on the sphere, just as in five spacetime dimensions \cite{Fuentealba:2021yvo,Fuentealba:2022yqt}. A similar situation is encountered in electromagnetism \cite{Henneaux:2019yqq,Fuentealba:2023huv}.

\section{Asymptotic conditions - final version \label{Asymptotia1}}

If we collect the restrictions on the gauge parameters derived above, including those on the gauge parameters entering the expressions of the leading terms in the asymptotic expansions of the fields, we get as asymptotic conditions 
\begin{eqnarray}
h_{rr} && =\frac{1}{r^{2}}\lambda^{(0)}+\frac{1}{r^{3}}\xbar h_{rr}+\frac{1}{r^{4}}h_{rr}^{(2)}+o(r^{-4})\,,\label{eq:hrr2}\\
h_{rA} && = \frac{1}{r}\lambda_{A}^{(1)}+\frac{1}{r^{2}}\xbar h_{rA}+\frac{1}{r^{3}}h_{rA}^{(2)}+o(r^{-3})\,,\label{eq:hrA2}\\
h_{AB} && =r\theta_{AB}^{(0)}+\theta_{AB}^{(1)}+\frac{1}{r}\xbar h_{AB}+\frac{1}{r^{2}}h_{AB}^{(2)}+o(r^{-2})\,,\label{eq:hAB2}
\end{eqnarray}
and
\begin{eqnarray}
\pi^{rr} && =r^{2}\kappa_{(0)}^{rr}+r\kappa_{(1)}^{rr}+\xbar\pi^{rr}+\frac{1}{r}\pi_{(2)}^{rr}+o(r^{-1})\,,\label{eq:pirr2}\\
\pi^{rA} && =r\kappa_{(0)}^{rA}+\kappa_{(1)}^{rA}+\frac{1}{r}\xbar\pi^{rA}+\frac{1}{r^{2}}\pi_{(2)}^{rA}+o(r^{-2})\,,\label{eq:pirA2}\\
\pi^{AB} && =\kappa_{(0)}^{AB}+\frac{1}{r}\kappa_{(1)}^{AB}+\frac{1}{r^{2}}\xbar\pi^{AB}+\frac{1}{r^{3}}\pi_{(2)}^{AB}+o(r^{-3})\,.\label{eq:piAB2}
\end{eqnarray}
where the  coefficients of the leading terms are now given by
\begin{eqnarray}
\lambda^{(0)} && =-2F^{(0)}\,,\\
\lambda_{A}^{(1)} && =-\xbar D_{A}F^{(0)}\,,\\
\theta_{AB}^{(0)} && =\xbar D_{A}\xbar D_{B}U+\xbar g_{AB}U\,,\\
\theta_{AB}^{(1)} && =2\left(\xbar D_{A}\xbar D_{B}F^{(0)}+\xbar g_{AB}F^{(0)}\right)\,, \\
\kappa_{(0)}^{rr} && =\sqrt{\xbar g}\xbar\triangle V\,,\qquad \kappa_{(0)}^{rA}  =\sqrt{\xbar g}\xbar D^{A}V\,,\\
\kappa_{(1)}^{rr} && =-8\sqrt{\xbar g}N^{(0)}\,, \qquad \kappa_{(1)}^{rA}  =2\sqrt{\xbar g}\xbar D^{A}N^{(0)}\,,
\end{eqnarray}
and
\begin{eqnarray}
\kappa_{(0)}^{AB} && =\sqrt{\xbar g}\left(\xbar g^{AB}\xbar\triangle V-\xbar D^{A}\xbar D^{B}V\right)\,,\\
\kappa_{(1)}^{AB} && =\sqrt{\xbar g}\left(\frac{5}{4}\xbar g^{AB}\xbar\triangle N^{(0)}-\xbar D^{A}\xbar D^{B}N^{(0)}\right)\,.
\end{eqnarray}

The functions on the sphere appearing in these asymptotic expansions are restricted by the conditions
\be \left(\xbar\triangle+4\right)N^{(0)}=0
\ee
and
\begin{equation}
\mathcal{G}=\mathcal{O}\left(r^{-2}\right)\,,\quad\mathcal{G}_{r}=\mathcal{O}\left(r^{-2}\right)\,,\quad\mathcal{G}_{A}=\mathcal{O}\left(r^{-1}\right)\,.
\end{equation}

With these boundary conditions, the symplectic form is finite.

\section{Gauge transformations}
\label{Sec:GTFormAnd Algebra}

\subsection{Action of the gauge transformations on the asymptotic fields}

The gauge transformations that preserve these asymptotic conditions are
\begin{eqnarray}
\epsilon && =T+\frac{1}{r}T^{(1)}+\frac{1}{r^{2}}T^{(2)}+o(r^{-2})\,, \label{Eq:GTReduced1}\\
\epsilon^{r} && =W+\frac{1}{r}W^{(1)}+\frac{1}{r^{2}}W^{(2)}+o(r^{-2})\,,\label{Eq:GTReduced2}\\
\epsilon^{A} && =\frac{1}{r}D^{A}W+\frac{1}{r^{2}}D^{A}W^{(1)}+\frac{1}{r^{3}}I_{(2)}^{A}+o(r^{-3})\,,\label{Eq:GTReduced3}
\end{eqnarray}
where the functions $T$, $W$,  $W^{(1)}$,   $T^{(2)}$, $W^{(2)}$ and $I_{(2)}^{A}$ are arbitrary functions 
on the 4-sphere at spatial infinity and $\left(\xbar\triangle+4\right)T^{(1)}=0$.  The generators of these transformations are finite.  They  define proper gauge transformations when $T= W= \tilde{T}^{(2)} = \tilde{W}^{(2)}  = 0$, with $\tilde{T}^{(2)} =\left(\xbar\triangle+4\right)T^{(2)} $ and $\tilde{W}^{(2)}  =\xbar D_{A}I^{(2)A}-\xbar\triangle W^{(2)}$.

We shall call the gauge transformations parametrized by $T$ and $W$ ``leading gauge transformations'' or even ``leading supertranslations''.  At the same time, the gauge transformations parametrized by $\tilde{T}^{(2)}$ and $ \tilde{W}^{(2)}$ will be called ``subleading gauge transformations'' or ``subleading supertranslations''.

One easily verifies that
the leading orders in the fall-off of the fields transform
as
\begin{eqnarray}
\delta\lambda^{(0)} && =-2W^{(1)}\,, \qquad
\delta\xbar h_{rr}  =-4W^{(2)}\,,\\
\delta\lambda_{A}^{(0)} && =0\,, \qquad
\delta\lambda_{A}^{(1)}  =-\xbar D_{A}W^{(1)}\,,\\
\delta\xbar h_{rA} && =- 2\xbar D_{A}W^{(2)} \,,\\
\delta\theta_{AB}^{(0)} && = 2 \xbar D_{A} \xbar D_{B} W+2\xbar g_{AB}W\,,\\
\delta\theta_{AB}^{(1)} && =2 \xbar D_{A} \xbar D_{B} W^{(1)}+2\xbar g_{AB}W^{(1)}\,,\\
\delta\xbar h_{AB} && =\xbar D_{A}I_{B}^{(2)}+\xbar D_{B}I_{A}^{(2)}+2\xbar g_{AB}W^{(2)}\,, \\
\delta\kappa_{(0)}^{rr} && =-\sqrt{\xbar g}\xbar\triangle T\,, \qquad
\delta\kappa_{(1)}^{rr}  =8\sqrt{\xbar g}T^{(1)}\,,\\
\delta\xbar\pi^{rr} && =\sqrt{\xbar g}\left(8T^{(2)}-\xbar\triangle T^{(2)}\right)\,, \\
\delta\kappa_{(0)}^{rA} && =-\sqrt{\xbar g}\xbar D^{A}T\,, \qquad
\delta\kappa_{(1)}^{rA}  =-2\sqrt{\xbar g}\xbar D^{A}T^{(1)}\,,\\
\delta\xbar\pi^{rA} && =-3\sqrt{\xbar g}\xbar D^{A}T^{(2)}\,, \\
\delta\kappa_{(0)}^{AB} && =-\sqrt{\xbar g}\left(\xbar g^{AB}\xbar\triangle T-\xbar D^{A}\xbar D^{B}T\right)\,,\\
\delta\kappa_{(1)}^{AB} && =-\sqrt{\xbar g}\left(\frac{5}{4}\xbar g^{AB}\xbar\triangle T^{(1)}-\xbar D^{A}\xbar D^{B}T^{(1)}\right)\,,\\
\delta\xbar\pi^{AB} && =-\sqrt{\xbar g}\left(\xbar g^{AB}\xbar\triangle T^{(2)}-\xbar D^{A}\xbar D^{B}T^{(2)}\right)\,.
\end{eqnarray}

Hence, we have in particular the following transformation rules for $U$, $V$, 
\be
\delta U  =2 W\,,\qquad\delta V=-T\,,
\ee
as well as 
$\delta F^{(0)}  =W^{(1)} $ and $\delta N^{(0)}=-T^{(1)}$.

\subsection{Algebra of the gauge generators}

As we have just shown, the variables $U$ and $V$ transform simply by shifts under leading supertranslations.  Since $U$ and $V$ are the charges of the subleading supertranslations, it follows that the abelian algebra of leading and subleading supertranslations is centrally extended in its canonical realization.

One finds indeed
\be
 \{G[T], G[\tilde{W}^{(2)}] \} = -2\oint d^{4}x\sqrt{\xbar g}\tilde{W}^{(2)}T\,, \quad \{G[W], G[\tilde{T}^{(2)}] \} = 4\oint d^{4}x\sqrt{\xbar g}\tilde{T}^{(2)}W\,,
\ee
where the $l=0$ mode in the first term and the $l = 1$ mode in the second one are projected out by definition of $\tilde{T}^{(2)}$ and $\tilde{W}^{(2)}$, respectively. 

Because the leading and subleading supertranslations form conjugate pairs, they can be decoupled from the Lorentz algebra by using the general argument of \cite{Fuentealba:2022xsz,Fuentealba:2023hzq} (see  \ref{App:Poinc}).

\section{Conclusions and comments}
\label{Sec:Conclusions}
In this paper, we have studied the asymptotic structure of linearized gravity in six spacetime dimensions. We have paid a special attention on the supertranslations and have shown that these are parametrized by four functions of the angles, as in five dimensions.  The supertranslations are $\mathcal O(1)$ and $\mathcal O(r^{-2})$ gauge transformations with gauge parameters taking a specific form ensuring finiteness of the symplectic structure and of the charges.  These parameters depend on two functions of the angles at order  $\mathcal O(1)$ and on two other functions of the angles at order  $\mathcal O(r^{-2})$.    In each case,  one of these functions parametrizes the temporal diffeomorphisms, the other parametrizes the radial diffeomorphisms (the angular components of the supertranslations are not independent).

The $\mathcal O(r^{-1})$ gauge transformations, which might have been anticipated to define also non trivial supertranslations, turn out to be proper gauge transformations with vanishing charges. 

Furthermore, we have shown that the supertranslation charges provide a central extension of the abelian algebra of the supertranslations. Leading ($\mathcal O(1)$) supertranslations and relevant subleading ($\mathcal O(r^{-2})$) ones form canonically conjugate pairs.

\subsection{Higher spacetime dimensions}

The pattern in higher spacetime dimensions $D$ is exactly the same: the non trivial supertranslations remain parametrized by four functions of the angles: two functions describing $\mathcal O(1)$ gauge transformations and two functions describing $\mathcal O(r^{-(D-4)})$ gauge transformations.  The intermediate orders $\mathcal O(r^{-k)})$ ($k = 1, \cdots , D-5$) define proper gauge transformations.    

Leading  and relevant subleading supertranslations form canonically conjugate pairs.
Note that for $D=4$, the two orders coincide since $\mathcal O(r^{-(D-4)}) = \mathcal O(1)$: leading and subleading supertranslations are the same and are  just the standard $D=4$ supertranslations.  This was one of our motivations for introducing logarithmic supertranslations in \cite{Fuentealba:2022xsz}, to bring in canonical conjugates to the standard supertranslations\footnote{In four spacetime dimensions, there is the additional need to impose parity conditions, which cuts the number of functions describing the supertranslations in two. Taking this into account, standard supertranslations and logarithmic supertranslations are described by two functions of the angles (and not four).}.

We have verified that this pattern holds in all spacetime dimensions $\geq 5$. We give here the explicit formulas for the gauge parameters and the corresponding charge-generators.  

The gauge transformations verifying all the finiteness requirements have the asymptotic expansion
\begin{eqnarray}
\epsilon && = T^{(0)} + \sum_{m=1}^{D-5}\frac{T^{(m)}}{r^{m}}+ \frac{T^{(D-4)}}{r^{D-4}}+ \mathcal{O}\left(r^{-(D-3)}\right)\,,\\
\epsilon^{r} && = W^{(0)}+ \sum_{m=0}^{D-5}\frac{W^{(m)}}{r^{m}}+\frac{W^{(D-4)}}{r^{D-4}}+ \mathcal{O}\left(r^{-(D-3)}\right)\,,\\
\epsilon^{A} && = \frac{\xbar D^A W_{(0)}}{r}+ \sum_{m=1}^{D-5}\frac{I_{(m)}^{A}}{r^{m+1}}+ \frac{I_{(D-4)}^{A}}{r^{D-3}}+\mathcal{O}\left(r^{-(D-2)}\right)\, , 
\end{eqnarray}
with 
\begin{equation}
\left(\xbar\triangle+D-2\right)T^{(m)}= 0\,,\quad \xbar D_A I^{A}_{(m)}-\xbar \triangle W^{(m)}=0  \qquad (0<m<D-4)\, .
\end{equation}

The asymptotic conditions for the fields $h_{ij}$ and $\pi^{ij}$ are obtained by acting with such gauge transformations on the following configurations
 \begin{eqnarray}
h_{rr} & =\frac{1}{r^{D-3}}\xbar h_{rr}+\frac{1}{r^{D-2}}h_{rr}^{(2)}+o(r^{-(D-2)})\,,\\
h_{rA} & =\frac{1}{r^{D-4}}\xbar h_{rA}+\frac{1}{r^{D-3}}h_{rA}^{(2)}+o(r^{-(D-3)})\,,\\
h_{AB} & =\frac{1}{r^{D-5}}\xbar h_{AB}+\frac{1}{r^{D-4}}h_{AB}^{(2)}+o(r^{-(D-4)})\,,
\end{eqnarray}
and
\begin{eqnarray}
\pi^{rr} && =\xbar\pi^{rr}+\frac{1}{r}\pi_{(2)}^{rr}+o(r^{-1})\,,\\
\pi^{rA} && =\frac{1}{r}\xbar\pi^{rA}+\frac{1}{r^{2}}\pi_{(2)}^{rA}+o(r^{-2})\,,\\
\pi^{AB} && =\frac{1}{r^{2}}\xbar\pi^{AB}+\frac{1}{r^{3}}\pi_{(2)}^{AB}+o(r^{-3})\,,
\end{eqnarray}
This will generate terms of a specific form that generically start at $(D-4)$ powers of $r$ higher than the leading order in the above equations (except for $h_{rr}$ and $h_{rA}$ where it is $(D-5)$  powers of $r$).  

The canonical generator of the asymptotic symmetries is given by
\begin{equation}
G=\int d^{D-1}x\left(\epsilon\mathcal{G}+\epsilon^{i}\mathcal{G}_{i}\right)+Q[\epsilon,\epsilon^{i}]\,,
\end{equation}
where the surface integral $Q[\epsilon,\epsilon^{i}] \equiv Q$ reads
\begin{eqnarray}
Q && =\oint d^{D-2}x\Big[\sqrt{\xbar g}T^{(0)}\left((D-2)\xbar h_{rr}+2\xbar D^{A}\xbar h_{rA}+(D-4)\xbar h\right)+W^{(0)}\left(\xbar\pi^{rr}-\xbar\pi\right) \nonumber \\
&& \qquad \qquad - (D-4)\sqrt{\xbar g}\tilde{T}^{(D-4)}U-2\sqrt{\xbar g}\,\tilde{W}^{(D-4)}V\Big]\,,\label{Eq:Surface7}
\end{eqnarray}
where 
\begin{eqnarray}
\tilde{T}^{(D-4)} & =\left(\xbar\triangle+D-2\right)T^{(D-4)}\,,\\
\tilde{W}^{(D-4)} & =\xbar D^{A}I_{A}^{(D-4)}-\xbar\triangle\,W^{(D-4)}\,,
\end{eqnarray}
and where $U$ and $V$ are the parameters $W^{(0)}$ and $T^{(0)}$ entering the gauge transformation in the asymptotic form of the fields.  Note that on the $(D-2)$-sphere, the operator $\xbar\triangle+D-2$ projects out  the $l=1$ spherical harmonic component.

The expression for the charges shows that the intermediate orders $T^{(m)}$ and  $W^{(m)}$ ($1 \leq m \leq D-5$) define proper gauge transformations, since they do not appear in the surface integral (\ref{Eq:Surface7}).  The only relevant subleading supertranslations are those at order $(D-4)$ with $\tilde{W}^{(D-4)} \not=0$.  We call them again for short ``subleading supertranslations''.  

It also follows that leading and subleading supertranslation charges form canonically conjugate pairs,  since the charges $V$ and $U$ of the subleading supertranslations are shifted by the leading supertranslations.

\subsection{Comparing with null infinity}

As we mentioned in the introduction, the study of the asymptotic symmetries of gravity at null infinity has been performed in \cite{Kapec:2015vwa} in even spacetime dimensions, also within the linearized context.  This work explores the implications of one specific type of supertranslations for the soft theorems, namely, in our language, the leading supertranslations with parity conditions imposed (as can be read off from their matching conditions between past null infinity and future null infinity). These supertranslations depend on one single function of the angles and are the supertranslations that reduce to the ordinary supertranslations in $D=4$. In terms of the parametrization (\ref{Eq:GTReduced1})-(\ref{Eq:GTReduced3}), the only non-zero functions are $T$ and $W$, with $T = T^{even}$ and $W = W^{odd}$.  It would be interesting to understand the implications of the other supertranslations.

This requires a careful matching of the asymptotic fields between spatial and null infinity, extending to gravity what has been done in \cite{Henneaux:2019yqq} for electromagnetism. Work along these lines is in progress.

\vspace{1cm}

\section*{Acknowledgments}
We thank C\'edric Troessaert for important discussions.  This work was partially supported by a Marina Solvay Fellowship (O.F.) and by  FNRS-Belgium (conventions FRFC PDRT.1025.14 and IISN 4.4503.15), as well as by funds from the Solvay Family.

\vspace{1cm}

\appendix

\section{Spatial infinity: other approaches ($D=4$)}
\label{App:Comparison}

Different definitions of symmetries lead to different symmetry groups.  The definitions might differ in the ``object'' that the symmetry should preserve.

A (classical) theory is defined by an action principle and, in case the spatial sections have a boundary, also by boundary conditions - which are asymptotic conditions if the ``boundary'' is at infinity.

Our definition of a symmetry of a theory is that it should preserve this structure.  More precisely, a transformation of the dynamical fields is a symmetry if (i) it preserves the boundary conditions; and (ii) it leaves the action invariant up to surface contributions at the time boundaries (as in classical mechanics - but no contributions at the spatial boundaries).  This second condition implies in particular that the symplectic $2$-form (encoded in the action) is strictly invariant under the transformation, and not merely invariant up to a possibly non-vanishing surface term. One can then construct the canonical generators of the symmetry  (moment map).

\subsection{Twisted parity conditions -- BMS group}
With the boundary  conditions proposed in \cite{Henneaux:2018hdj,Henneaux:2019yax}, the above definition of a symmetry yields the BMS group as asymptotic symmetry group at spatial infinity.  The boundary conditions of \cite{Henneaux:2018hdj,Henneaux:2019yax} differ from the strict parity conditions that the authors of \cite{Regge:1974zd} imposed on the leading orders of the fields, by a $\mathcal O(1)$ diffeomorphism taking a specific form compatible with Lorentz invariance.  One says that the parity conditions are twisted by a diffeomorphism.  [A different twist in the parity conditions was considered in \cite{Henneaux:2018cst}.  It also leads to the BMS symmetry.  Although perfectly fine from the Hamiltonian point of view developed on spacelike hypersurfaces, this alternative set of boundary conditions does not eliminate the leading log-singularity at null infinity.  One might argue that this is a problem of null infinity, not of the theory.]

\subsection{Strict parity conditions or reduction to Poincar\'e}

At spatial infinity, the charges are strictly conserved.  In particular, the BMS charges are conserved.  This means that the twist in the parity conditions is constant.  By making a definite diffeomorphism, one can thus undo the twist and assume that the strict parity conditions of \cite{Regge:1974zd} hold.
This reduces the BMS group to the Poincar\'e group.  Although it was not formulated in terms of a reduction of a larger group, the emergence of the Poincar\'e symmetry at spatial infinity was exhibited long ago in value of the ADM energy, momentum and angular momentum.

One could even argue that since all the Poincar\'e charges are conserved, one could go to the rest frame of the system once and for all, where the group is  further reduced to $\mathbb{R} \times SO(3)$ (which is not an invariant subgroup, in much the same way as the Poincar\'e group is not an invariant subgroup of the BMS group).

However, the strict conservation of the BMS charges only holds for a hypothetical observer sitting at an infinite distance from the source in a non cosmological background.  Even though such observers do not exist, asymptotic methods are useful, but one must take into account that they are approximate.  In particular, there might be fluxes that change the charges (see e.g., \cite{Misner:1973prb} chapter 19).

Imagine an observer sitting at a very large distance from a gravitational source, which detects gravitational waves from the source.  Before the waves hit the observer, the above asymptotic expansion of the metric and conjugate momentum can be used.  One can compute in particular from the asymptotic values of the fields  definite ``initial'' values of the ADM energy, momentum and angular momentum.  The asymptotic expansion is also valid after the wave has passed, leading to ``final'' values of the ADM energy, momentum and angular momentum which are in general different from the initial ones. The difference is what has been carried away by the wave. If the momentum has changed, one ceases to be in the rest frame of the observed system (although the hypothetical observer at infinity would remain of course in the rest frame of the full system including the wave).  This is one reason why it is not advisable to reduce the Poincar\'e group to the little group defined by the initial values of the conserved quantities.
The situation is quite clear on physical grounds. 

The ADM mass, momentum and linear momentum are not the only BMS quantities changed by the passage of the wave.  Pure supertranslation charges will also change. Indeed, even if the metric and its conjugate momentum obey initially the strict (leading order) parity conditions of \cite{Regge:1974zd}, they will generically not do so after the wave has passed, unless the wave is very carefully parity-tuned.  A non-trivial twist, i.e., non-trivial values of $U$ and $V$ will be generated.  This will be invisible to the non-existent (never hit) observer at infinity, but it corresponds to a real effect for real observers at a non-infinite distance: it is the memory effect \cite{Strominger:2014pwa},\cite{Fuentealba:2023syb}.  For that reason, it would be equally unadvisable to reduce the BMS group to the Poincar\'e group by imposing strict parity conditions on the grounds that the wave never reaches spatial infinity.

The problem of digging a Poincar\'e subalgebra within the BMS algebra in order to extract a supertranslation-free angular momentum has actually been 
bypassed in recent studies, which have shown that non-linear redefinitions effectively achivieve this task.  The proper Hamiltonian context where  all the relevant quantities have a well-defined action through the Poisson bracket is given in \cite{Fuentealba:2022xsz}.

\subsection{$Spi$ approach \cite{Ashtekar:1978zz,Ashtekar:2023wfn,Ashtekar:2023zul}}

In a very insightful work \cite{Ashtekar:1978zz} (see also \cite{Ashtekar:2023wfn,Ashtekar:2023zul}), which has inspired many developments, a rich geometrical structure is attached to spatial infinity.  An asymptotic symmetry is then defined as a transformation preserving this structure.  This leads to the $Spi$ group \cite{Ashtekar:1978zz}, which is similar but different from the BMS group: it is also the semi-direct product of the Poincar\'e group with an abelian group of ``supertranslations'', but the $Spi$ supertranslations do not coincide with the BMS ones.  The difference in the resulting symmetry groups is not contradictory but is a consequence of the difference in what is required to be preserved in the definition of a symmetry.  

Since the role of the gravitational action is not as central in \cite{Ashtekar:1978zz} as it is in our approach, the construction of a well-defined moment map (with ``integrable'', finite generators) for the $Spi$ group is less direct.  To our knowledge the moment map has in fact not been built for all the $Spi$ generators.  Interesting considerations on the construction of the $Spi$ charges for some of the $Spi$ supertranslations can be found in \cite{Compere:2011ve,Troessaert:2017jcm}.

\section{Poincar\'e transformations}
\label{App:Poinc}
We have carefully checked Poincar\'e invariance of the theory, which amounts to verify that the Poincar\'e transformations leave the action invariant (up to boundary terms on the initial and final slices but without surface contributions at spatial infinity).  To achieve invariance, it is necessary to supplement the original form of the Poincar\'e transformations (\ref{eq:dh-Poincare})-(\ref{eq:dp-Poincare}) by  ``correcting gauge transformations'', which we have explicitly worked out.  

{We also found it necessary to impose the condition $\lambda^{(0)A}= 0$.  This is  as in four spacetime dimensions, where this condition is imposed to ensure invariance of the kinetic term under boosts, i.e.,  integrability of the boost charges (more information on this point in  \cite{Henneaux:2019yax}).}

Because the improved Poincar\'e transformations preserve the action,  they preserve in particular the symplectic form and define therefore canonical transformations.  Their charge-generators $P_{\xi,\xi^{i}}$, where  $(\xi,\xi^{i})$ are the components of the spacetime vector fields defining the infinitesimal Poincar\'e transformations (rotations, boosts and translations), can be determined by the standard methods and are the sum of a bulk term involving the energy-momentum tensor of the linearized theory and surface terms.

The computations are rather long and cumbersome and will not be reproduced here since they are not very illuminating (and in direct line of what can be found in the literature \cite{Fuentealba:2022yqt,Fuentealba:2020ghw}). 
The brackets of the Poincar\'e generators give the expected result,
namely,
\begin{equation}
\{P_{\xi_{1},\xi_{1}^{i}},P_{\xi_{2},\xi_{2}^{i}}\}=P_{\hat{\xi},\hat{\xi}^{i}}\,,
\end{equation}
where
\begin{eqnarray}
\hat{\xi}^{i} & =\xi_{1}^{j}\partial_{j}\xi_{2}^{i}+g^{ij}\xi_{1}\partial_{j}\xi_{2}-(1\leftrightarrow2)\,,\\
\hat{\xi} & =\xi_{1}^{i}\partial_{i}\xi_{2}-(1\leftrightarrow2)\,.
\end{eqnarray}

The Poisson brackets of the canonical generators of the gauge symmetries with the Poincar\'e generators are given by
\begin{equation}
\{G_{\epsilon},P_{\xi,\xi^{i}}\}=G_{\hat{\epsilon}}\,,
\end{equation}
with
\begin{eqnarray}
\hat{T} && =-Y^{A}\partial_{A}T+5bW+\partial_{A}b\xbar D^{A}W+b\xbar\triangle W\,,\\
\hat{W} && =-Y^{A}\partial_{A}W+bT\,,\\
\hat{\tilde{T}}^{(2)} && =-Y^{A}\partial_{A}\tilde{T}^{(2)}+\frac{5}{2}b\tilde{W}^{(2)}+\frac{1}{2}\partial_{A}b\xbar D^{A}\tilde{W}^{(2)}+\frac{1}{2}b\xbar\triangle\tilde{W}^{(2)}\,,\\
\hat{\tilde{W}}^{(2)} && =-Y^{A}\partial_{A}\tilde{W}^{(2)}+2b\tilde{T}^{(2)}\,.
\end{eqnarray}
(The factors of $2$ appearing in the expression of $\hat{\tilde{T}}^{(2)}$ and $\hat{\tilde{W}}^{(2)}$ as compared with the expression of $\hat{T}$ and $\hat{W}$ come  from a different normalization and can be absorbed through the redefinition ${\tilde{T}}^{(2)} \rightarrow {\tilde{T}}'^{(2)} = 2 {\tilde{T}}^{(2)}$ so that $\hat{\tilde{T}}'^{(2)}  =-Y^{A}\partial_{A}\tilde{T}'^{(2)}+ 5b\tilde{W}^{(2)}+\partial_{A}b\xbar D^{A}\tilde{W}^{(2)}+b\xbar\triangle\tilde{W}^{(2)} $ and $\hat{\tilde{W}}^{(2)}  =-Y^{A}\partial_{A}\tilde{W}^{(2)}+b\tilde{T}'^{(2)}$).  The algebra does not contain non-linear terms as in  \cite{Fuentealba:2022yqt} ($D=5$) but this is of course because we are dealing with linearized gravity.  

The non-vanishing brackets $\{G_{\epsilon},P_{\xi,\xi^{i}}\}$ show that the generators $G_{\epsilon}$ transform in a non-trivial representation of the Poincar\'e group.  We recall that as one goes to the full Einstein theory, the Poincar\'e transformations become improper gauge symmetries with generators given on-shell by surface integrals.  The translations combine with the (leading) supertranslations by ``filling the holes $l=0$ (time translations) and $l=1$ (space translations)'' \cite{Fuentealba:2020ghw}.  The brackets $\{G_{\epsilon},P_{\xi,\xi^{i}}\}$ encode also the transformation rule of the translations under Lorentz transformations. 

Finally, because the pure supertranslations (leading and subleading) form canonically conjugate pairs, one can decouple them (and only them, not the ordinary translations of course) from the Poincar\'e algebra by non-linearly redefining the Lorentz generators \cite{Fuentealba:2023hzq}. 

\newpage

\noindent
{\bf {\large References}}

\end{document}